\documentclass[aps,prb,10pt,twocolumn,superscriptaddress,floatfix,showpacs]{revtex4-1}
\usepackage{graphicx}
\usepackage{epsfig}% Include figure files
\usepackage{bm}% bold math
\usepackage{amssymb,amsmath}
\usepackage{url,hyperref}
\usepackage{color}

\usepackage{lineno}%{stepnumber=1}

\newcommand \med[1] {\langle{#1}\rangle}
\newcommand \ze[1] {|{#1}\rangle}
\newcommand \ez[1] {\langle{#1}|}

\begin{document}

%\linenumbers

\title{Application of the S=1 underscreened Anderson lattice model to Kondo uranium and
neptunium compounds.}

\author{Christopher Thomas}
\affiliation{Instituto de F\'{\i}sica, Universidade Federal do Rio
Grande do Sul, 91501-970 Porto Alegre, Brazil}
\affiliation{Institut N\'{e}el, CNRS-UJF, BP 166, 38042 Grenoble Cedex 9, France}
\author{Acirete S. da Rosa Sim\~oes}
\affiliation{Instituto de F\'{\i}sica, Universidade Federal do Rio
Grande do Sul, 91501-970 Porto Alegre, Brazil}
\author{J. R. Iglesias}
\affiliation{Instituto de F\'{\i}sica, Universidade Federal do Rio
Grande do Sul, 91501-970 Porto Alegre, Brazil}
\author{C. Lacroix}
\affiliation{Institut N\'{e}el, CNRS-UJF, BP 166, 38042 Grenoble Cedex 9, France}
\author{N. B. Perkins}
\affiliation{Department of Physics, University of Wisconsin-Madison,
Madison, WI 53706, USA}
\author{B. Coqblin}
\affiliation{Laboratoire de Physique des Solides, Universit\'{e}
Paris-Sud, UMR-8502 CNRS, 91405 Orsay, France}

\date{\today}

\begin{abstract}
%\nolinenumbers 
Magnetic properties of uranium and neptunium compounds showing the coexistence of Kondo screening
effect and ferromagnetic order are investigated within the Anderson lattice Hamiltonian with a two-fold
degenerate $f$-level in each site, corresponding to $5f^2$ electronic configuration with $S=1$ spins. A
derivation of the Schrieffer-Wolff transformation is presented and the resulting Hamiltonian has an effective
$f$-band term, in addition to the regular exchange Kondo interaction between the $S=1$ $f$-spins and the $s=1/2$
spins of the conduction electrons. The obtained effective Kondo lattice model can describe both the Kondo regime
and a weak delocalization of $5f$-electron. Within this model we compute the Kondo and Curie temperatures as a
function of model parameters, namely the Kondo exchange interaction constant $J_K$, the magnetic intersite
exchange interaction $J_H$ and the effective $f$-bandwidth. We deduce, therefore, a phase diagram of the model
which  yields the coexistence of Kondo effect and ferromagnetic ordering and also  accounts for the pressure
dependence of the Curie temperature of uranium compounds such as UTe.

\end{abstract}


\maketitle

\section{\label{sec:intro}INTRODUCTION}

In cerium, ytterbium, uranium and other anomalous rare-earth and actinide
compounds, the interplay between Kondo effect and magnetism leads to the
formation of various interesting phenomena which still attract large attention.
\cite{Hewson,Coqblin,Stewart} Both effects depend strongly on the
hybridization between  $f$ and  conduction electrons, which in turn
significantly influence the level of localization of the $f$-electrons. The
extent of the localization is sensitive to various external parameters, such
as temperature and pressure, but most importantly to the spatial extension of
the orbitals. Actually, in the case of cerium compounds, 4$f$-electrons are
usually well localized, while in the case of uranium and other actinide
compounds, 5$f$-electrons can be either localized or itinerant, or in-between,
depending on the studied system. Moreover, in the case of 5$f$-electrons, the
experimental data do not clearly distinguish between a localized 5$f^n$
configuration and a mixed-valence regime. Thus, there is always a challenge to
decide which is the best framework for discussing actinide materials. The
nature of the electronic structure of actinide metals has been studied since a
long time and extensively
reviewed.~\cite{IglesiasJ,Cooper,cooper98,cooper99,Zwicknagl1,Zwicknagl2,Moore,Santini}

The difference between the $4f$- and $5f$-electrons leads to
different magnetic properties of rare-earth and actinide
compounds.~\cite{BC} In the case of cerium Kondo compounds, a
competition between the Kondo effect on each Ce atom and the
magnetic ordering of the Ce magnetic moments through the RKKY
interaction has been successfully described by the so-called
"Doniach diagram". \cite{Doniach} In this diagram, both the N\'eel
temperature $T_N$ (or the Curie temperature $T_C$) and the Kondo
temperature $T_K$ are obtained as functions of the Kondo exchange
interaction constant, $J_K$. In most of the Cerium compounds,  the
magnetic ordering temperatures, $T_N$ or $T_C$, are rather low,
typically of order 5 to 10 K. With increasing pressure (i.e. with
increasing $J_K$) ordering temperatures pass through a maximum and
then tend to zero at the Quantum Critical Point, above which the
systems are non magnetic heavy fermion ones.

The situation in actinide compounds is more complex, and it is now
established that some of them exhibit a coexistence of magnetic
order and Kondo effect. Indeed, this phenomenon was  observed in
several uranium compounds, like
UTe,\cite{Schoenes1,Schoenes2,Schoenes3}
UCu$_{0.9}$Sb$_2$,\cite{Buko} or UCo$_{0.5}$Sb$_2$\cite{Tran} in
which a ferromagnetic order with large Curie temperatures (equal,
respectively, to $T_C = 102$ K, 113 K and 64.5 K) and a logarithmic
Kondo-type decrease of the resistivity above $T_C$ has
 been experimentally detected. A similar behavior has been observed in the neptunium compounds, NpNiSi$_2$
\cite{Colineau} and Np$_2$PdGa$_3$,\cite{Tran2} with Curie temperatures equal to, respectively, $T_C =$51.5 K
and 62.5 K.

The origin of this fundamentally different behaviors lies in the
fact that 5$f$-electrons are generally less localized than the
4$f$-electrons and have tendency to partial delocalization. In the
latter case, 5$f$-electrons have dual, localized and delocalized,
nature  which leads to reduction of magnetic moments. Both phenomena
have been experimentally detected. For example, magnetic moments
observed in UTe are substantially smaller than the free-ion values
for either the $5f^2$ or the $5f^3$ configurations. \cite{Schoenes3}
On the same side, in the series of uranium monochalcogenides, US
lies closest to the itinerant side for the $5f$-electrons, USe is in
the middle and the $5f$-electrons are more localized in UTe, as
evidenced by magnetization measurements.\cite{Schoenes3} Moreover,
the Curie temperature of UTe is passing through a maximum and is
then decreasing with applied pressure, which is interpreted as a
weak delocalization of the $5f$-electrons under
pressure.\cite{Schoenes3,Cooper} The dual nature of the
$5f$-electrons, assuming two localized $5f$-electrons and one
delocalized electron, has been also studied by Zwicknagl {\it et
al.} \cite{Zwicknagl1,Zwicknagl2} who have accounted, by using band
calculations, for the mass enhancement factor experimentally
observed in some uranium compounds.

Another important difference between magnetic cerium and uranium
compounds lies in the value of the $5f$-electron spins $S$ which is
always larger than 1/2 in uranium systems (To avoid any confusion,
in the following $S$ designs always the spin of the $f$-localized
electrons and $s$ the spin of the conduction electrons). For spins
$S$ larger than 1/2, Kondo effect is more complex and depends on the
number of screening channels: the underscreened Kondo impurity
problem and more generally the multichannel impurity problem have
been studied extensively \cite{Nozieres, LeHur, Parcollet, Florens,
Coleman1} and   solved exactly by the Bethe Ansatz method.
\cite{Schlottmann} It has been clearly established that the large
spin $S$ of the $f$-electrons can be completely screened at $T=0$ by
spins of  conduction electrons only if the number of screening
channels $n$ is  equal to $2S$. If this is not the case, the problem
is more complicated as shown by Nozieres and Blandin.
\cite{Nozieres}

Here we are interested by the case of the underscreened Kondo effect, which
occurs when $n$ is smaller than $2S$; in this case, in contrast to the regular
Kondo impurity case, the spin is only partially screened at low temperatures,
and this leads  to a reduced effective spin $S_{\text{eff}} =S - n/2$. For a
spin $S=1$, this happens if there is only one active screening channel, i.e.,
when only one conduction band is present near the Fermi level. Indeed, for the
systems studied here, this is an oversimplification: generally all screening
channels are not equivalent, thus, it is natural to consider that only one
channel is coupled more strongly to the local spin. This channel will dominate
the behavior, but other screening channels might play a role at lower
temperatures, resulting in a two-stage Kondo screening with two Kondo
temperatures.\cite{Pustilnik} Very recently, the underscreened impurity
problem has been widely discussed in relation with experiments performed on
quantum dots  coupled to ferromagnetic leads,\cite{Weymann} and in molecular
quantum dots (so called molecular transistors).\cite{Roch}

In a concentrated Kondo system, these reduced effective spins $S_{\text{eff}}
=S - n/2$ interact ferro- or antiferromagnetically through RKKY exchange
interaction, leading to magnetic ordering of the reduced moments. Thus
coexistence of magnetic ordering and Kondo effect is expected to occur more
easily in the underscreened case than in the standard $S=1/2$ Kondo lattice.

The first attempt to describe the coexistence of ferromagnetism and Kondo
effect in uranium compounds has been performed with the help of an
underscreened Kondo lattice (UKL) model which considered localized $f$-spins
$S$=1 to describe a 5$f^2$ configuration of uranium
ions.\cite{Perkins2,Perkins} This model describes the Kondo interaction,
$J_K$, between localized $S$=1 spins and $s=1/2$  spins of conduction
electrons, and an inter-site ferromagnetic exchange interaction between the
$f$-spins, $J_H$. The mean-field treatment of the UKL model gives an analog of
the Doniach phase diagram, and qualitatively accounts for the coexistence of
ferromagnetism and Kondo effect. However, this model is based on the
assumption of localized 5$f$-electrons. Of course, this assumption makes a
constraint on systems which can be described by the UKL, because, as we
already discussed, many metallic actinides do not have fully localized
electrons. To improve the model for these compounds, one needs to include in
the UKL a possibility for 5$f$-electron delocalization. This is the main goal
of the present study.

The article is organized as follows: we start by considering, in section
\ref{sec:SW}, the underscreened Anderson lattice (UAL) model, in which the
charge transfer is present from the beginning, and we transform it by using
the Schrieffer-Wolff transformation\cite{SW} for $n^f_{\text{tot}}=2$, which
allows for an effective $f$-band term, in addition to the  exchange Kondo
interaction between the $S=1$ $f$-spins and the $s=1/2$ spins of the
conduction electrons. Then, in section \ref{sec:MF} we present the mean field
treatment of the model and in section \ref{sec:RC} we compute the Curie and
Kondo temperatures as a function of the different parameters and in particular
of the $f$-bandwidth. We obtain  new phase diagrams which could account for
the pressure dependence of uranium systems such as UTe compound.

\section{\label{sec:SW} THE S=1 SCHRIEFFER-WOLFF TRANSFORMATION.}

The standard model to describe  the physics of the heavy fermion
compounds is the periodic Anderson lattice model whose  Hamiltonian
can be written as:

%\begin{eqnarray}
\begin{align}
\label{Hander}
H = H_{c} + H_{V} + H_{f}~.
\end{align}
%\end{eqnarray}
The first term describes a conduction $c$-electron band:

%\begin{eqnarray}
\begin{align}
\label{Hcond}
H_{c}= \sum_{{\bf k}\sigma} \epsilon_{\bf k}c^{\dagger}_{{\bf k}\sigma}c_{{\bf k}\sigma}~,
\end{align}
%\end{eqnarray}
where $c^{\dagger}_{{\bf k}\sigma}$  creates a conduction
quasiparticle  with spin $\sigma$ and momentum ${\bf k}$, and
$\epsilon_{\bf k}$ is the energy of conduction electrons. The term
$H_f$ includes all local energy terms of the $f$-electrons and it is
given by:

%\begin{eqnarray}
\begin{align}
\label{Hcorr}
%\begin{array}{l}
H_{f}&=\sum_{i\sigma\alpha}E^{f}n^{f}_{i\alpha\sigma}+\sum_i\Big[U(n_{i1\uparrow}^f n_{i1\downarrow}^f
+ n_{i2\uparrow}^f n_{i2\downarrow}^f) \nonumber\\
&\quad+U'(n_{i1\uparrow}^fn_{i2\downarrow}^f+n_{i1\downarrow}^fn_{i2\uparrow}^f)\nonumber\\
&\quad+(U'-J)(n_{i1\uparrow}^fn_{i2\uparrow}^f+n_{i1\downarrow}^fn_{i2\downarrow}^f)\nonumber\\
&\quad-J(f^{\dagger}_{i1\uparrow}f_{i1\downarrow}f^{\dagger}_{i2\downarrow}f_{i2\uparrow}
+ h. c.)\Big]~,
%\end{array}
\end{align}
%\end{eqnarray}

where $E^f$ is the energy of the two-fold degenerate $f$-level and
$n^{f}_{i\sigma\alpha}$ is the number operator for $f$-electrons on
lattice site $i$, orbital $\alpha$ and spin $\sigma$, $U$ and $U'$
are  the Coulomb repulsions among electrons in the same and in the
different orbitals, respectively, and $J$ is the Hund's coupling
constant. For $2$ electrons per site, the ground state of $H_f$ is
the triplet state with $S=1$. We assume here that this triplet state
is much lower in energy than the singlet states. This is achieved if
$U'-J$ is much smaller than $U'+J$ and than $U$. We study the SW
transformation in this limit, assuming that 2 electrons on the same
site can only be coupled in the $S=1$ state. Both subsystems,
localized $f$-electrons and conduction band, are coupled via a
hybridization term,

%\begin{eqnarray}
\begin{align}
\label{Hhyb}
H_{V}= \sum_{i{\bf k}\sigma\alpha} (V_{\bf k\alpha}e^{i{\bf k}\cdot{\bf R}_i}c^{\dagger}_{{\bf k}\sigma}f_{i\alpha\sigma} +
V^{*}_{\bf k\alpha}e^{-i{\bf k}\cdot{\bf R}_i}f^{\dagger}_{i\alpha\sigma}c_{{\bf k}\sigma}),
\end{align}
%\end{eqnarray}
where $f_{i\sigma\alpha}^{\dagger}$ and $f_{i\sigma\alpha}$ are creation and annihilation operators for
$f$-electrons, carrying spin and orbital indexes,  $\sigma$ and  $\alpha~(\alpha=1,2)$, respectively.

In the Schrieffer-Wolff transformation, the $f-c$ hybridization term is eliminated by a canonical
transformation. Thus, using the classical method explained in Ref. \onlinecite{SW}, we start the
procedure by writing the Hamiltonian as:

%\begin{eqnarray}
\begin{align}
H=H_0+H_V~,
\end{align}
%\end{eqnarray}

where $H_0=H_c+H_f$. Then, we look at the scattering of an initial
state $\ze{a}$ to a final state $\ze{b}$, through an intermediate
state $\ze{c}$, where the states $\ze{a}$, $\ze{b}$ and $\ze{c}$ are
eigenstates of $H_0$ and $E_a$, $E_b$ and $E_c$ are their
eigenvalues, respectively. The SW transformation consists in
replacing the $H_V$ term of the Hamiltonian by an effective
interaction which is of second order in the hybridization parameter
$V$ of the starting Hamiltonian $H_V$. The detailed description of
the calculations can be found in Ref. \onlinecite {Hewson} and
\onlinecite {SW}. The resulting effective Hamiltonian is given by:

%\begin{eqnarray}
\begin{align}
H\simeq H_0+\tilde{H}~,
\end{align}
%\end{eqnarray}
where:

%\begin{eqnarray}
\begin{align}
\ez{b}\tilde{H}\ze{a}=\frac{1}{2}\sum_{c}\ez{b}H_V\ze{c}\ez{c}H_V\ze{a}% \nonumber \\
(\frac{1}{E_a-E_c}+\frac{1}{E_b-E_c})~.
\end{align}
%\end{eqnarray}

The SW transformation has been initially performed for the case of one $4f$-electron in the $4f^1$
configuration. Here, we will present a derivation of the SW transformation in the case of a $5f^2$
configuration, corresponding to a $f$-spin $S=1$, and to the so-called  "Underscreened Kondo Lattice" model
where the $S=1$ spins are coupled to a non-degenerate conduction band.

We study the SW transformation for 2 $f$-electrons per site and allowing
fluctuations of the number of $f$-electrons between 1 and 2. Several
interactions are generated by the SW transformation.\cite{ChrisBuzios} Below
we present the derivation of the most relevant terms, leading to both local
and intersite effective interactions, namely the Kondo interaction and the
effective hopping of $f$-electrons.

\subsection{\label{sec:sub:LEI} The local effective interaction}

First, we derive the $s-f$ exchange Hamiltonian for $S=1$ localized $f$-spins.
The corresponding eigenstates of $H_0$ are, therefore, given by:

\begin{align}
\ze{a}&=c_{{\bf k}\sigma'}^{\dag}f_{i1\sigma}^{\dag}f_{i2\sigma}^{\dag}\ze{0}~, \nonumber \\
\ze{b}&=\frac{c_{{\bf k}\sigma}^{\dag}}{\sqrt{2}}\left(f_{i1\uparrow}^{\dag}f_{i2\downarrow}^{\dag}
+f_{i1\downarrow}^{\dag}f_{i2\uparrow}^{\dag}\right)\ze{0}~,\nonumber \\
&=\frac{1}{\sqrt{2}}\sum_{\sigma'}c_{{\bf k} \sigma}^{\dag}f_{i1\sigma'}^{\dag}f_{i2\bar{\sigma}'}^{\dag}\ze{0}~,\nonumber \\
\ze{c}&=c_{{\bf k}\sigma}^{\dag}c_{{\bf k'}\sigma'}^{\dag}f_{i\alpha\sigma''}^{\dag}\ze{0}~,
\end{align}
and the corresponding eigenvalues are

%\begin{eqnarray}
\begin{align}
E_{a}&= E_{b} = U'-J+2E^f+\epsilon_{{\bf k}}~, \nonumber \\
E_{c}&=\epsilon_{{\bf k}}+\epsilon_{{\bf k'}}+E^f~.
\end{align}
%\end{eqnarray}

The SW transformation leads to the standard Kondo-like $s-f$ exchange Hamiltonian, but here
with $f$-spins $S=1$, and it is given by:

%\begin{eqnarray}
\begin{align}
H_{K}&=\frac{1}{2}\sum_{i{\bf k}{\bf k'}}J_{{\bf k},{\bf k'}}\Big[c_{{\bf k'}\uparrow}^{\dag}c_{{\bf k}\downarrow}S_i^-
+c_{{\bf k'}\downarrow}^{\dag}c_{{\bf k}\uparrow}S_i^+\nonumber \\
&\quad+(c_{{\bf k'}\uparrow}^{\dag}c_{{\bf k}\uparrow}-c_{{\bf k'}\downarrow}^{\dag}c_{{\bf k}\downarrow})S_{i}^z\Big]
\end{align}
%\end{eqnarray}
where the different components of the spin $S=1$ read:

%\begin{eqnarray}
\begin{align}
S^+_i&=n_{i1}^ff_{i2\uparrow}^{\dag}f_{i2\downarrow}
+f_{i1\uparrow}^{\dag}f_{i1\downarrow}n_{i2}^f~,\nonumber \\
S^-_i&=n_{i1}^ff_{i2\downarrow}^{\dag}f_{i2\uparrow}
+f_{i1\downarrow}^{\dag}f_{i1\uparrow}n_{i2}^f~,\nonumber \\
S^z_i&=n_{i1\uparrow}^fn_{i2\uparrow}^f
-n_{i1\downarrow}^fn_{i2\downarrow}^f~,
\end{align}
%\end{eqnarray}
and the corresponding exchange integral is:

%\begin{eqnarray}
\begin{align}
J_{\bf k,\bf k'}&=-V_{{\bf k}\alpha}^* V_{{\bf k'}\alpha} e^{i({\bf k- k'})\cdot{\bf R}_i}\nonumber \\
&\quad\times\Big(\frac{1}{U'-J+E^f-\epsilon_{\bf k'}}
+\frac{1}{U'-J+E^f-\epsilon_{\bf k}}\Big)~,
\end{align}
%\end{eqnarray}

This exchange integral $J_{\bf k,\bf k'}$ can be easily simplified, because $\epsilon_{\bf k}$ can be restricted
to values very close to the Fermi energy. Then, the difference between ${\bf k}$ and ${\bf k'}$ can be neglected
in both the values of $\epsilon_{\bf k}$ and of $V_{\bf k\alpha}$. We also assume that the mixing parameter does
not depend on the  orbital index $\alpha$. Thus, $J_{\bf k,\bf k'}$ can be approximated in the following by:

%\begin{eqnarray}
\begin{align}
\label{eqJk}
J_{\bf k,\bf k'}\approx-\frac{2|V_{k_F}|^2}{U'-J+E^f-\mu}\equiv J_K~,
\end{align}
%\end{eqnarray}
where $\mu$ is the Fermi level and $V_{k_F}$ is the value of $V$ at the Fermi level. Eq. \eqref{eqJk} gives the
definition of the Kondo exchange interaction, $J_K$ that we will use in the following. It is also interesting to
notice that the Kondo effect is large when the energy $U'-J+E^f$ is very close to the Fermi energy. In fact the
denominator in Eq. \eqref{eqJk} is the energy difference  between the ground state energy
$U'-J+2E^f+\epsilon_{\bf k}$ of two $f$-electrons  in $\ze{a}$ and $\ze{b}$ states  and the energy of the
intermediate state  $\ze{c}$ with one $f$-electron: $E^f+\epsilon_{\bf k}+\epsilon_{\bf k'}$.

\subsection{\label{sec:sub:II} The intersite effective interaction}

Among the different terms emerging from the SW transformation we consider in detail those that correspond to a
non-local interaction involving 2 different sites $i$ and $j$. In this case, the relevant initial and final
states $\ze{a}$ and $\ze{b}$ are two-sites states with a total occupation of three $f$-electrons, allowing
charge fluctuations between sites $i$ and $j$. Consequently, an effective $f$ bandwidth is obtained in
the $2^{nd}$ order in $V_{\bf k}$. Thus, we derive an effective band Hamiltonian, $H_W$, as the sum of three
terms which arise from the Schrieffer-Wolff transformation and where the sum over $i$ and $j$ refers to
different sites:

%\begin{eqnarray}
\begin{align}
H_{W}= H_{W1} + H_{W2} + H_{W3}~,
\end{align}
%\end{eqnarray}

where:

%\begin{eqnarray}
\begin{align}\label{hw1}
H_{W1}&=-\sum_{{\bf k}\alpha\sigma ij}\frac{|V_{\bf k}|^2e^{i{\bf k}\cdot({\bf R}_i-{\bf R}_j)}}{U'-J+E^f-\epsilon_{\bf k}} \nonumber \\
&\quad\times(f^{\dagger}_{j\alpha\sigma} f^{\dagger}_{i1\sigma}
f^{\dagger}_{i2\sigma} f_{j2\sigma} f_{j1\sigma} f_{i\alpha\sigma}
\nonumber \\ &\quad- f^{\dagger}_{j\alpha\sigma} f^{\dagger}_{i1\sigma}
f^{\dagger}_{i2\sigma} f_{j2\sigma} f_{j1\sigma}
f_{i\bar{\alpha}\sigma} +h.c.)~,
\end{align}
%\end{eqnarray}

%\begin{eqnarray}
\begin{align}\label{hw2}
H_{W2}&=-\frac{1}{2}\sum_{{\bf k}\alpha\sigma \sigma'ij}\frac{|V_{\bf k}|^2e^{i{\bf k}\cdot({\bf R}_i-{\bf R}_j)}}{U'-J+E^f-\epsilon_{\bf k}}
 \nonumber \\
&\quad\times(f^{\dagger}_{j\alpha \bar{\sigma}} f^{\dagger}_{i1\sigma}
f^{\dagger}_{i2\sigma} f_{j2\bar{\sigma}'} f_{j1\sigma'}
f_{i\alpha\sigma} \nonumber \\ &\quad- f^{\dagger}_{j\alpha\bar{\sigma}}
f^{\dagger}_{i1\sigma} f^{\dagger}_{i2\sigma} f_{j2\bar{\sigma'}}
f_{j1\sigma'} f_{i\bar{\alpha}\sigma} +h.c.)~,
\end{align}
%\end{eqnarray}

%\begin{eqnarray}
\begin{align}\label{hw3}
H_{W3}&=-\frac{1}{4}\sum_{{\bf k}\alpha \sigma \sigma' \sigma''ij}
\frac{|V_{\bf k}|^2e^{i{\bf k}\cdot({\bf R}_i-{\bf R}_j)}}{U'-J+E^f-\epsilon_{\bf k}} \nonumber \\
&\quad\times(f^{\dagger}_{j\alpha\sigma} f^{\dagger}_{i1\sigma'}
f^{\dagger}_{i2\bar{\sigma'}} f_{j2\bar{\sigma}"} f_{j1\sigma''}
f_{i\alpha\sigma} \nonumber \\ &\quad- f^{\dagger}_{j\alpha\sigma}
f^{\dagger}_{i1\sigma'} f^{\dagger}_{i2\bar{\sigma'}}
f_{j2\bar{\sigma''}} f_{j1\sigma''} f_{i\bar{\alpha}\sigma}
+h.c.)~.
\end{align}
%\end{eqnarray}

These three terms can be simplified in the mean field approximation by introducing average occupation numbers
$\med{n_{i\alpha\sigma}^f}$. For simplicity, we will  neglect all interorbital transfer terms. With these
assumptions, we can see that the $H_W$ terms can be considered as effective $f$-hopping terms between $i$ and
$j$ sites with a spin dependent hopping (see next section). If the ${\bf k}$ dependence of $V_{\bf k}$ is
neglected, as is often assumed, one can deduce from Eqs. \eqref{hw1}, \eqref{hw2} and \eqref{hw3} that there is
no intersite hopping. Thus, the ${\bf k}$ dependence of $V_{\bf k}$ is at the origin of the effective
$f$-bandwidth and has to be taken into account. This ${\bf k}$ dependence is due to non-local hybridization
between $f$ and $c$ electrons, which has no influence on the Kondo interaction, but has to be taken into account
for intersite interactions. In the following we write the coefficient which enters in the expression of $H_W$
as:

\begin{align}
\frac{|V_{\bf k}|^2}{U'-J+E^f-\epsilon_{\bf k}}\approx-\frac{J_K}{2}g({\bf k})~,
\end{align}
where the function $g({\bf k})$ includes the ${\bf k}$-dependence of $|V_{\bf k}|^2$, while the ${\bf
k}$-dependence of $\epsilon_{\bf k}$ is not essential here, but could be also included.

Finally, the resulting transformed Hamiltonian contains two terms: $H_{K}$, that gives the Kondo exchange
interaction for spin $S=1$ and the important and new term $H_W$, which can be considered as an effective band
term for the $5f$-electrons.

Thus, in addition to the Kondo interaction, we have derived here a term which
leads to a finite $f$-bandwidth; this newly introduced term will be shown as
giving a better description of uranium and actinide compounds where the
$5f$-electrons are less localized than the $4f$-electrons in rare earth
compounds.

Besides this effective band term, there are other intersite interactions which
lead to RKKY exchange, but they arise only to $4^{th}$ order in hybridization
$V_{\bf k}$. We do not compute all these terms, but, instead, we will
introduce them phenomenologically as an additional intersite exchange
parameter of the model, $J_H$.

\section{\label{sec:MF} THE MEAN FIELD APPROACH.}

Combining all terms obtained in the preceding section, we can write the new
effective Hamiltonian as:

%\begin{eqnarray}
\begin{align}
{\mathcal H}= H_{c} +  H_{K} + H_{W} + \frac{1}{2} J_H \sum_{ij}{\bf S}_i{\bf S}_j~.
\end{align}
%\end{eqnarray}

The Heisenberg interaction  $J_H$ is considered here as a ferromagnetic exchange between nearest neighbors only.
In fact RKKY interactions are long range and oscillating, but since our aim is to study the coexistence of
ferromagnetism and Kondo effect, we consider only ferromagnetic interactions. The mean field approach has been
 previously described in Ref. \onlinecite{Perkins}. For the Kondo part $H_{K}$, it is based on a generalization of a
 functional integration approach described by Yoshimori and Sakurai for the single impurity case
\cite{Yoshimori} and by Lacroix and Cyrot for the $S=1/2$ Kondo lattice case.\cite{Lacroix,Coleman2}

Here we introduce the following mean field parameters: the average occupation numbers $\med{n_{i\alpha\sigma}^f}$,
$\med{n_{i\sigma}^c}$ and the "Kondo" parameter $\med{\lambda_{i\alpha\sigma}}=\med{f^{\dagger}_{i\sigma}
c_{i\alpha\sigma}}$. We restrict ourselves to uniform solutions in which all these quantities are site and orbital independent,
i.e., $\med{n_{j\alpha\sigma}^f}=\med{n_{i\alpha\sigma}^f}=n_{\sigma}^f$, $\med{n_{i\sigma}^c}=n_{\sigma}^c$
 and $\med{\lambda_{i\alpha\sigma}}=\lambda_{\sigma}$.

Thus, $H_W$ can be written as:

%\begin{eqnarray}
\begin{align}
H_W=\sum_{{\bf k}\sigma\alpha}\Gamma_{{\bf k}\sigma}f_{{\bf
k}\alpha\sigma}^{\dagger} f_{{\bf k}\alpha\sigma}~,
\end{align}
%\end{eqnarray}

The effective band dispersion depends on spin but not on the orbital indexes.
Then:

\begin{align}
\Gamma_{{\bf k}\sigma}=\Gamma_{\sigma}g({\bf k}) \nonumber \\
\end{align}
with
\begin{align}
\Gamma_{\sigma}&=- \frac{J_K}{2} \Big[(n_{\sigma}^f)^2+\frac{1}{2}n_{\sigma}^f
n_{\bar{\sigma}}^f
+\frac{1}{4}(n_{\bar{\sigma}}^f)^2\Big]~, \nonumber \\
\end{align}
and where $g({\bf k})$ is the dispersion relation for the $f$-band. We assume,
for simplicity, that the $f$-band dispersion is similar to the conduction
electron dispersion, i. e., $g({\bf k})=P\epsilon_{\bf k}+P'$. We also assume
that the $f$-band should be narrower than the conduction band, so $P < 1$.
Parameter $P'$ can be included in the local energy $E^f$, and $P$ is a
multiplicative coefficient that can be absorbed in the definition of
$\Gamma_{\sigma}$. Then we have:

\begin{align}\label{eqgks}
\Gamma_{{\bf k}\sigma}=\Gamma_{\sigma} g({\bf k}) = A_{\sigma} \epsilon_{\bf k}
\end{align}
with
\begin{align}\label{eqas}
A_{\sigma}&=- \frac{J_K}{2} P \Big[(n_{\sigma}^f)^2+\frac{1}{2}n_{\sigma}^f
n_{\bar{\sigma}}^f
+\frac{1}{4}(n_{\bar{\sigma}}^f)^2\Big]~,
\end{align}
%\end{eqnarray}

\begin{figure}[ht]
\begin{center}
\includegraphics[width=0.4\textwidth]{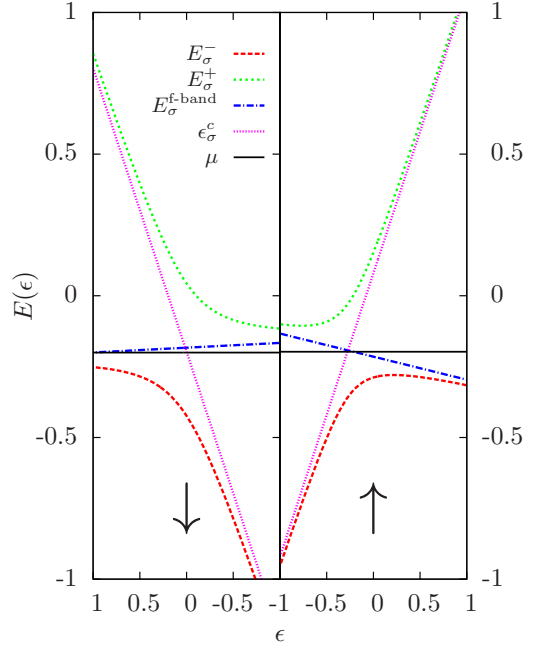}
\end{center}
\caption{Band structure obtained in mean field approximation, for $P=0.12$, $J_K=0.53$, $J_H=-0.01$, $n^c=0.8$ and
$n^f_{\text{tot}}=2$.} \label{figdos}
\end{figure}

\begin{figure}[ht]
\begin{center}
\includegraphics[width=0.4\textwidth]{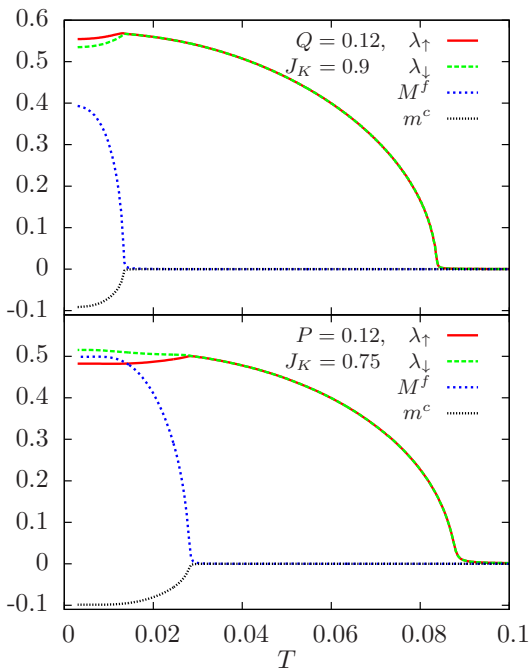}
\end{center}
\caption{Magnetization of $f$-electrons $M^f$, magnetization of $c$-electrons
$m^c$, and Kondo parameters $\lambda_{\sigma}$ for both spin directions for
cases (b) and (c) (see the text for definitions), with $J_H=-0.01$, $n^c=0.8$
and $n^f_{\text{tot}}=2$.} \label{figq012p20}
\end{figure}

The total Hamiltonian can be now written in the mean field approach
as follows:

%\begin{eqnarray}
\begin{align}
\label{hff}
{\mathcal H}&=\sum_{{\bf k}\sigma}\epsilon_{{\bf k}\sigma}^c n_{{\bf k}\sigma}^c+\sum_{i\sigma\alpha}E_{0\sigma}^f n_{i\alpha \sigma}^f \nonumber \\
&\quad+\sum_{{\bf k}\sigma\alpha}\Lambda_{\sigma}
(c_{{\bf k}\sigma}^{\dagger}f_{{\bf k}\alpha\sigma}+h.c.) \nonumber \\
&\quad+\sum_{{\bf k}\sigma}A_{{\bf k}\sigma}f_{{\bf k}\alpha\sigma}^{\dagger}f_{{\bf k}\alpha\sigma} + C~,
\end{align}
%\end{eqnarray}

where we have :

%\begin{eqnarray}
\begin{align}
E_{0\sigma}^{f}&=E^f+U'n_{\bar{\sigma}}^f+(U'-J)n_{\sigma}^f
+J_K\sigma m^c \nonumber \\
&\quad-\frac{J_K}{8}(\lambda_{\uparrow} + \lambda_{\downarrow})^2+J_Hz\sigma M^f~,\\
%\end{eqnarray}
%\begin{eqnarray}
\epsilon_{{\bf k}\sigma}^c&=\epsilon_{\bf k}+\Delta_{\sigma}~,\qquad \Delta_{\sigma}=J_K\sigma M^f~,\\
%\end{eqnarray}
%\begin{eqnarray}
\Lambda_{\sigma}&=-\frac{J_K}{4} (\lambda_{\sigma} + \lambda_{\bar{\sigma}})~,\\
%\end{eqnarray}
%\begin{eqnarray}
C&=-2U'N n_{\uparrow}^f n_{\downarrow}^f%\nonumber \\
-(U'-J)N[(n_{\uparrow}^f)^2 + (n_{\downarrow}^f)^2]\nonumber \\
&\quad+\frac{J_K}{2}N (\lambda_{\uparrow} + \lambda_{\downarrow})^2-\frac{J_H}{2}zN (M^f)^2 \nonumber \\
&\quad-J_K N m^c M^f~,
\end{align}
%\end{eqnarray}
with $\sigma=\pm\frac{1}{2}$, $M^f=n_{\uparrow}^f-n_{\downarrow}^f$ and
$m^c=\frac{1}{2}(n_{\uparrow}^c-n_{\downarrow}^c)$.

The diagonalization of the Hamiltonian gives one pure $f$-band, $E^{f}_{{\bf
k}\sigma}$ given by :

%\begin{eqnarray}
\begin{align}
E^{f}_{{\bf k}\sigma}=E_{0\sigma}^{f}+ A_{\sigma}\epsilon_{{\bf k}}~,
\end{align}
%\end{eqnarray}
and two hybridized bands $E^{\pm}_{{\bf k} \sigma}$ given by:

\begin{align}
E^{\pm}_{{\bf k} \sigma}=\frac{1}{2}[\epsilon_{{\bf k}}(1+A_{\sigma})+E_{0\sigma}^{f}+ \Delta_{\sigma} \pm
S_{{\bf k}\sigma}]~,
\end{align}
with

\begin{align}
S_{{\bf k} \sigma}&=E^{+}_{{\bf k} \sigma} - E^{-}_{{\bf k} \sigma} \nonumber \\
&=\sqrt{[\epsilon_{{\bf k}}(1 - A_{\sigma})-(E_{0\sigma}^{f}-\Delta_{\sigma})]^2 +8(\Lambda_{\sigma})^2}~.
\end{align}

On Fig. \ref{figdos}, we present a typical band structure resulting from the three bands $E^{f}_{{\bf k}\sigma}$
and $E^{\pm}_{{\bf k}\sigma}$ for $J_K=0.53$ and $J_H=-0.01$. In all figures presented here, $z=6$, where $z$ is
the number of nearest neighbors in a simple cubic lattice. In our calculations, the values of  $J_K$ and  $J_H$
are defined in units of the half bandwidth $D$ of the conduction band. One can see on Fig. \ref{figdos} the
important effect of the finite $f$-bandwidth: the band structure is very different from that without any
$f$-bandwidth used in Ref. \onlinecite{Perkins}.

\section{\label{sec:RC} RESULTS AND CONCLUSIONS}

In this section, we present numerical results obtained from this model, using
the general method described in detail in Ref. \onlinecite{Perkins}: we
derive the Green functions and we calculate self-consistently  the
magnetization $M^f$ for the $f$-electrons, the magnetization $m^c$ for the
conduction electrons and the two spin-dependent $\lambda_{\sigma}$ parameters
which describe the Kondo effect, by imposing constraints on the total number
of $f$-electrons and conduction electrons, $n^f_{\text{tot}}=2$, and
$n^c_{\text{tot}}=n^c$, respectively. Having solved the self-consistent
equations, we study various properties of the model. The Curie and Kondo
temperatures are defined, within this mean field approach, as the temperatures
at which respectively the magnetizations or the $\lambda_{\sigma}$ parameters
tend to zero.

%\begin{widetext}
\begin{figure*}[ht]
%\begin{center}
\includegraphics[width=0.8\textwidth]{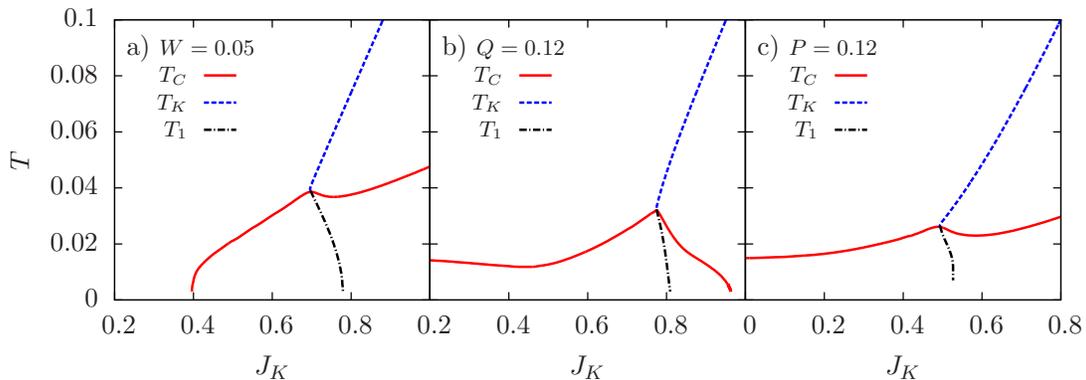}
%\end{center}
\caption{Curie temperature $T_C$, and Kondo temperature $T_K$, versus $J_K$
for the three cases (a), (b) and (c), with $J_H=-0.01$, $n^c=0.8$ and
$n^f_{\text{tot}}=2$. For cases (a), (b) and (c) $T_1$ is also shown (see the
text).} \label{figtctk}
\end{figure*}
%\end{widetext}

\begin{figure}[ht]
\begin{center}
\includegraphics[width=0.4\textwidth]{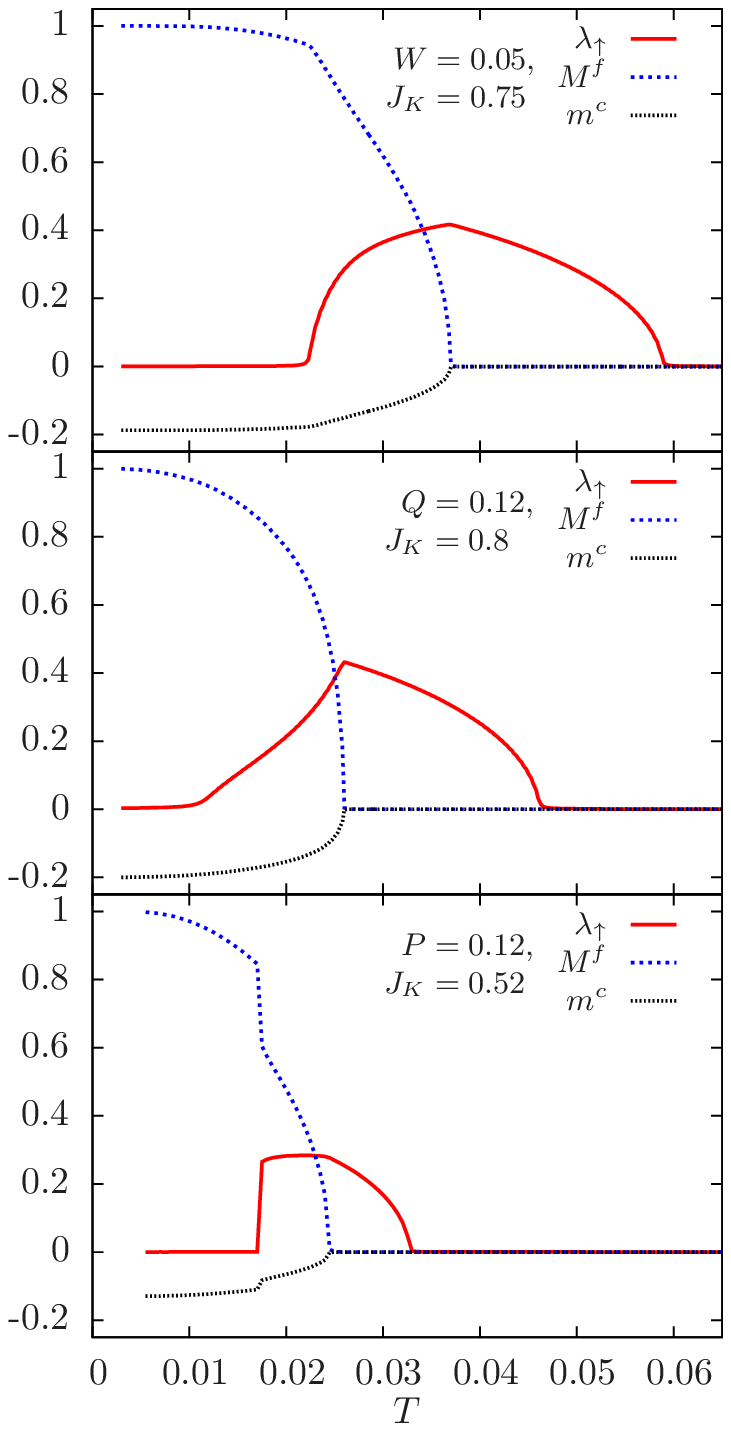}
\end{center}
\caption{Magnetization of $f$-electrons $M^f$, magnetization of $c$-electrons
$m^c$, and Kondo parameter $\lambda_{\uparrow}$ for cases (a), (b) and (c),
with $J_H=-0.01$, $n^c=0.8$ and $n^f_{\text{tot}}=2$. Here
$\lambda_{\downarrow}\approx\lambda_{\uparrow}$.} \label{fig4}
\end{figure}

As already mentioned in the previous section, the half $f$-bandwidth derived from the Schrieffer-Wolff
transformation is spin-dependent and it is given by $A_{\sigma}$, eq. \eqref{eqas}. $A_{\sigma}$ can be rewritten as:

%\begin{eqnarray}
\begin{align}
A_{\sigma}=- \frac{J_K P}{32}\Big[7+3(M^f)^2+6\sigma M^f\Big]~.
\end{align}
%\end{eqnarray}
indicating that the effective bandwidth and the magnetization are correlated. In fact it can be easily checked
that the bandwidth for $up$-spin increases with magnetization while it is the opposite for $down$-spin: this is
consistent with the double exchange process in the ferromagnetic phase, which favors itinerancy of the
conduction electrons with spin parallel to the localized moment, because of intra-atomic Hund's coupling.

Here, however, we would like to explore the parameter dependence of the effective bandwidth including different
possibilities for the relative variation of the Kondo coupling, $J_K$, and the $f$-bandwidth $W_f$ . In order to
do that, we  considered also  the following definitions of $W_f$:
\begin{itemize}
\item case (a): a constant bandwidth: $W_f=const$;
\item case (b): a bandwidth  $W_f$ proportional to the Kondo coupling constant:
$W_f=QJ_K$; in this way  we can take into account the effect of pressure on
both the bandwidth and the Kondo coupling, since both are sensitive to the
increase of hybridization under pressure.
\item case (c): a bandwidth directly obtained from the SW transformation. From
Eqs. \eqref{eqgks}-\eqref{eqas}, we get a spin dependent bandwidth: $W_f=2A_{\sigma}$.
\end{itemize}
In the following, all calculations are done assuming that the conduction band
$\epsilon_{\bf k}$ has a width $2D$ and that its density of states is constant
and equal to $1/2D$.

On Fig. \ref{figq012p20}, we present a plot of the temperature variation of the Kondo correlations
$\lambda_{\uparrow}$ and $\lambda_{\downarrow}$, and also the $f$ and $c$ magnetizations,  $M^f$ and $m^c$, for
cases (b) and (c). The parameters are $n^c=0.8$ and  $J_H=-0.01$. Upper plot is for case (b) with $Q=0.12$,
while lower plot is for case (c)  with $P=0.12$.

The two magnetization curves clearly show a second order magnetic phase transition at $T_C$, below which $m^c$
and $M^f$ are always antiparallel, as expected because $J_K$ is an antiferromagnetic coupling. At low
temperatures, Kondo effect and ferromagnetism coexist together, and, due to the breakdown of spin symmetry at
$T_C$, $\lambda_{\uparrow}$ and $\lambda_{\downarrow}$ become slightly different in the ferromagnetic phase. We
define the Kondo temperature as the temperature where $\lambda_{\uparrow}$ and $\lambda_{\downarrow}$ vanish. The fact
that the Kondo parameter vanishes at a particular temperature is a well known artefact of the mean field
approximation. Actually $T_K$ is a crossover temperature, associated with the onset of Kondo screening. In all
cases, the Kondo temperature $T_K$ is larger than the Curie temperature $T_C$ and we never found situations
where $T_K<T_C$: once ferromagnetism is established, Kondo effect does not appear below $T_C$; it is blocked by
the effective internal magnetic field.

To investigate the effect of the pressure on the Kondo  and Curie
temperatures, $T_K$ and $T_C$, respectively, we computed  these
characteristic temperatures for various values of $J_K$ for fixed
values of exchange interaction $J_H$ and number of conduction
electrons $n_c$. Figs. 3a,3b and 3c  are obtained for three
different characterizations of the f-bandwidth, cases (a), (b) and
(c), respectively. We notice that the temperatures $T_K$ and $T_C$
are obtained as the temperatures at which the mean field parameters
($f$ and $c$ magnetizations and the Kondo parameters $\lambda$)
vanish. In the three cases we note that the Kondo temperature $T_K$
becomes non-zero only above a critical value $J_{K}^{c}$ which
varies  from case to case. In all cases, once non-zero, the Kondo
temperature rapidly increases for larger values of $J_K$. The Curie
temperature, $T_C$, is non-zero above a given $J_K$ value in case
(a), below a given $J_K$ value for case (b) and is non-zero for all
studied values of $J_K$ for case (c). The reason for these different
behaviors is easy to understand. In case a) the f-bandwidth is
constant and  the system needs a finite value of $J_K$ to get
magnetic ordering because f-electrons are itinerant even at small
$J_K$. In case b) the f-bandwidth increases linearly with $J_K$, so,
for low values of $J_K$ the f-electrons are localized and they are
magnetic  even for $J_K=0$; thus as soon as $J_K$ is different from
zero, magnetic ordering occurs. When increasing $J_K$ the
f-bandwidth  also increases, and  magnetism is destroyed because of
the itinerant character of the f-electrons. Finally in case c)   the
f-bandwidth depends on both $J_K$ and magnetization, and the
dependence is different for up and down spin electrons;it can be
seen on figure 3 that this complex dependence of the bandwidth leads
to small variation of the Curie temperature, with a weak maximum.
However a crossing point, at which $T_C$ and $T_K$ are equal is
obtained in all cases.

Concerning the Kondo effect, in all cases (a), (b) and (c) a
peculiar behavior has been obtained for values of $J_K$ just above
this crossing point: at the temperature $T_1$ indicated on Fig.
\ref{figtctk}a), \ref{figtctk}b) and \ref{figtctk}c) the Kondo
parameters vanish, being non-zero only between $T_1$ and $T_K$. To
better understand this behavior, we have plotted on Fig. \ref{fig4},
$M^f$, $m^c$ and $\lambda_{\uparrow}$ for $n^c=0.8$, $J_H=-0.01$ but
for values of $J_K$ near the crossing of $T_C$ and $T_K$, i.e.,
$J_K=0.75$ for case (a), $J_K=0.8$ for case (b) and $J_K=0.52$ for
case (c). It appears clearly that, with decreasing temperature, the
Kondo effect occurs first, then there is a coexistence of Kondo
effect and ferromagnetism, and finally the Kondo effect disappears
to yield only a strong ferromagnetism at extremely low temperatures.
This behavior can be interpreted in the following way: Kondo effect
for a spin $S=1$ cannot be complete, as explained in the
introduction. Thus if exchange is large enough, the ordering of the
remaining $f$-moments occurs in the Kondo phase. However, at lower
temperature,  when these magnetic moments are large, they act as an
internal magnetic field that destroys the Kondo effect. It should be
pointed out that there is at present no experimental evidence in
favor or in contrast of such an effect in actinide compounds at very
low temperature.

Another interesting result that can be pointed out is the decrease of the
Curie temperature  for large $J_K$ above the intersection point particularly
in case (b), but also within a small range of value of $J_K$ in case (c). This
decrease can probably be considered as resulting from the "delocalization" of
the 5$f$-electrons. Let us also remark that $J_K$ increases with increasing
pressure and that the two Figs. \ref{figtctk}b) and \ref{figtctk}c) can give a
description of the experimentally observed variation of $T_C$ with pressure in
UTe compound, which is passing through a maximum and then decreasing with
applied pressure.\cite{Schoenes3,Cooper}

To summarize our paper, the present work improves the previous $S=1$ UKL model of Ref. \onlinecite{Perkins} by
explicitly including the effect of a weak delocalization of the $5f$-electrons. Within this improved model, we
have obtained new phenomena in the region where $T_C$ and $T_K$ are of the same order of magnitude: a possible
disappearance of the Kondo effect at low temperature, which is a direct consequence of the underscreened Kondo
effect, and a maximum of $T_C$ as a function of $J_K$. It is worth to note that, in our model, the
delocalization of the $5f$-electrons increases when $J_K$ increases, i.e. when pressure is applied; then the
magnetization decreases and in the same way the Curie temperature. Therefore, the change in the Curie
temperature at large $J_K$ is more influenced by delocalization than by competition between magnetism and Kondo
effect. This is a different result compared with the case of Cerium compounds, where the magnetization is
destroyed by Kondo effect, i.e. by the screening of the magnetic moment. In the underscreened $S=1$ Kondo
lattice, because Kondo screening can never be complete, the Kondo effect alone does not destroy ferromagnetism.

To conclude, we have shown that our model includes two effects which
are essential to describe the $5f$-electron compounds: the small
delocalization of the $5f$-electrons, and the $S=1$ spins found in
uranium or neptunium compounds. The first effect works against
magnetism, while the second one favors magnetism. The competition
between these two effects leads to complex phase diagrams which can
improve the description of some actinide compounds and explain in
particular the maximum of $T_C$ observed experimentally in UTe
compound with increasing pressure.

\begin{acknowledgments}
This work was partially financed by Brazilian agency CNPq.
\end{acknowledgments}

\end{document}